\documentstyle[12pt,epsfig]{article}
 1
 1
 1

\begin{document}
\bibliographystyle{unsrt}
%
\def\lta{\;\raisebox{-.5ex}{\rlap{$\sim$}} \raisebox{.5ex}{$<$}\;}
\def\gta{\;\raisebox{-.5ex}{\rlap{$\sim$}} \raisebox{.5ex}{$>$}\;}

%

\newcommand{\permille}{$^0 \!\!\!\: / \! _{00}\;$}
\newcommand{\GeV}{GeV}
 
\newcommand{\mt}{m_{t}}
\newcommand{\mtt}{m_{t}^2}
\newcommand{\mw}{M_{W}}
\newcommand{\mww}{M_{W}^{2}}
\newcommand{\md}{m_{d}}
\newcommand{\mb}{m_{b}}
\newcommand{\mbb}{m_{b}^2}
\newcommand{\mc}{m_{c}}
\newcommand{\mh}{m_{H}}
\newcommand{\mhh}{m_{H}^2}
\newcommand{\mz}{M_{Z}}
\newcommand{\mzz}{M_{Z}^{2}}

\newcommand{\lra}{\leftrightarrow}
 
\newcommand{\ie}{{\em i.e.}}
\def\Ww{{\mbox{\boldmath $W$}}}  
\def\B{{\mbox{\boldmath $B$}}}         
\def\nn{\noindent}

\newcommand{\sinsq}{\sin^2\theta}
\newcommand{\cossq}{\cos^2\theta}
\newcommand{\be}{\begin{equation}}
\newcommand{\ee}{\end{equation}}
\newcommand{\ba}{\begin{eqnarray}}
\newcommand{\ea}{\end{eqnarray}}
\newcommand{\eea}{\end{eqnarray}}

\newcommand{\nl}{\nonumber \\}
\newcommand{\eqn}[1]{Eq.(\ref{#1})}
\newcommand{\ibidem}{{\it ibidem\/},}
\newcommand{\into}{\;\;\to\;\;}
\newcommand{\wws}[2]{\langle #1 #2\rangle^{\star}}
\newcommand{\p}[1]{{\scriptstyle{\,(#1)}}}
\newcommand{\ru}[1]{\raisebox{-.2ex}{#1}}
\newcommand{\epem}{$e^{+} e^{-}\;$}
\newcommand{\tch}{$t\to c H\;$}
\newcommand{\tcz}{$t\to c Z\;$}
\newcommand{\tcg}{$t\to c g\;$}
\newcommand{\tchm}{$t\to c H$}
\newcommand{\tczm}{$t\to c Z$}
\newcommand{\tcgm}{$t\to c g$}
\newcommand{\tcfm}{$t\to c \gamma$}
\newcommand{\tcht}{t\to c H}
\newcommand{\tczt}{t\to c Z}
\newcommand{\tcgt}{t\to c g}
\newcommand{\tcft}{t\to c \gamma}

\newcommand{\tbwht}{t\to b W H}
\newcommand{\tbwzt}{t\to b W Z}
\newcommand{\tbwh}{$t\to b W H\;$}
\newcommand{\tbwz}{$t\to b W Z\;$}
\newcommand{\tbwhm}{$t\to b W H$}
\newcommand{\tbwzm}{$t\to b W Z$}
\newcommand{\Gt}{\Gamma(t\to b W)}

\begin{titlepage}
\rightline{ROME1-1207/98}
\rightline{May 1998}
\vskip 22pt 
\vfil

\noindent
\begin{center}
{\Large \bf A new evaluation of the $t\to c H \;$ decay width \\ 
in the standard model.
}
\end{center}
\bigskip

\begin{center}
{\large 
B.~Mele~$^{a,b},$
$\;\,$S.~Petrarca~$^{b,a} \;\,$
and $\;\,$A.~Soddu~$^b$ 
} \\

\bigskip\noindent
$^a$ INFN, Sezione di Roma 1, Rome, Italy\\
\noindent
$^b$ Rome University ``La Sapienza", Rome, Italy
\end{center}
\bigskip
\begin{center}
{\bf Abstract} \\
\end{center}
{\small 
We present a new calculation of the \tch decay rate in the standard
model. 
We find that the corresponding branching fraction 
is $B(\tcht)\simeq 10^{-13} \div 10^{-14}$
for $\mz \lta \mh \lta 2\mw$, that is about 6 orders of magnitude
less than previously evaluated, and usually quoted in the literature.
}
\vskip 29pt 
\vfil
\noindent
e-mail: \\
mele@roma1.infn.it, $\;$
petrarca@roma1.infn.it, $\;$
soddu@roma1.infn.it \\ 
\end{titlepage}

Rare decays of the top quark have been extensively
studied from a theoretical point of view, both within
and beyond the standard model (SM) 
\cite{realiu,realid,sonid,intov,intoh,intox}.
The one-loop flavor-changing transitions, \tcgm, \tcfm, \tcz and \tchm,
are particularly interesting,
since they are  sensitive to possible effects from new physics,
such as supersymmetry, an extended Higgs sector and heavier-fermion families.
In the SM, these processes are in general quite suppressed due to the
Glashow-Iliopoulos-Maiani (GIM) mechanism \cite{gimmy}, 
controlled by the light masses of
the $b, s, d$ quarks circulating in the loop. 
The corresponding branching fractions $B_i=\Gamma_i/\Gamma_{T}$ 
are further decreased 
by the large total decay width $\Gamma_{T}$ of the top quark.
The complete calculations of the one-loop
flavour-changing top decays have been performed
before the experimental discovery of the top quark 
in the paper by
Eilam, Hewett and Soni ~\cite{sonid} 
(also based on Eilam, Haeri and Soni ~\cite{soniu}),
where, therefore, the top mass has been taken as a parameter.
Assuming  $\mt=175$GeV (according to
 the most recent value  measured at the Tevatron
$\mt=(173.9\pm 5.2)$GeV \cite{massa}), 
the value of the
total width $\Gamma_{T}\simeq \Gt$ is $\Gamma_{T} \simeq 1.55$ GeV,
and one gets from ref.~\cite{sonid}
\be
B(\tcgt)\simeq 4\cdot 10^{-11}, \;\;\;\; \\
B(\tcft)\simeq 5\cdot 10^{-13}, \;\;\;\;  \\
B(\tczt)\simeq 1.3\cdot 10^{-13}.
\label{giust}
\ee
In the same ref.~\cite{sonid}, 
a much larger branching fraction  for
the decay \tch is presented as function 
 of the top and Higgs masses.
We show in  Fig. 1 the relevant Feynman graphs for this channel.
In particular,
 for $\mt\simeq 175$ GeV and 40 GeV$\lta \mh \lta 2\mw$,
the authors of ref.~\cite{sonid} obtain
\be
B(\tcht)\simeq 10^{-7} \div 10^{-8}.
\label{sbagl}
\ee
This result was computed through the analytical formulae
presented in ref.~\cite{soniu}
for the fourth-generation quark decay  $b'\to bH$, in a 
theoretical framework assuming four flavour families.
Such  relatively large values for $B(\tcht)$ look surprising,
since the topology of the Feynman graphs  for the
 different one-loop channels is similar,
and  a GIM suppression, governed by the down-type
quark masses, is acting in all the decays.
The authors of ref.~\cite{sonid}
argue that in the channel \tch  the GIM suppression is softened by 
the Higgs-boson coupling to the quark. This statement seems puzzling,
since in the \tch amplitude the relevant Yukawa coupling
is of the order of $\mt/\mw$, which hardly can give rise to an
enhancement of many orders of magnitude in the corresponding width.

Another hint pointing to a problem with the large values of 
the \tch rate given above comes from  the comparison
between the rates for \tcz and \tch and the corresponding
rates for the tree-level decays \tbwz and \tbwhm, when $\mh \simeq \mz$.
The latter can be considered a sort of lower-order {\it parent}
processes for the one-loop decays, as can be seen
in Fig.~2, where we show the relevant Feynman graphs. 
In fact, the Feynman graphs for \tcz and \tch can be obtained 
by recombining the final $b$ quark and $W$ into a $c$ quark
in the three-body decays \tbwz and \tbwhm, respectively, 
and by adding analogous
contributions where the $b$ quark is replaced by the $s$ and $d$ quarks.
The main point of this comparison is  that the depletion of the
\tch rate with respect to the parent \tbwh rate
is expected to be of the same
order of magnitude of the depletion  of \tcz with respect to \tbwzm,
for $\mh \simeq \mz$. Indeed, the GIM mechanism acts in a similar way
in the one-loop decays into $H$ and $Z$.

An accurate determination of the \tbwz and \tbwh decay rates
has been carried out in ref.~\cite{realiu}.
This paper also takes into account the $W$ and $Z$ finite-width 
effects that are 
crucial, since the actual value of $\mt$ is (or can be) just near the
$bWZ$ (or $bWH$) threshold. For $\mh \simeq \mz$, the two widths are 
comparable. In particular, for $\mt\simeq 175$GeV, one has \cite{realiu}
\be
B(\tbwzt) \simeq 6\cdot 10^{-7}   \; \; \; \;\; \; \; \;\;  
B(\tbwht) \simeq 3\cdot 10^{-7}.
\label{brtree}
\ee
Assuming $B(\tcht) \simeq 6\cdot 10^{-8}$ that is the value
taken from ref.~\cite{sonid} for $\mh \simeq \mz$,
 one then  gets for the ratio of the one-loop
and tree-level decay rates
\be
r_H \equiv \frac{B(\tcht)}{B(\tbwht)} \sim 0.2  
\label{rhiggs}
\ee
to be confronted with 
\be
r_Z \equiv \frac{B(\tczt)}{B(\tbwzt)} \sim 2\cdot 10^{-7}.
\label{rzeta}
\ee
On the other hand, it seems natural to connect the values of $r_H$ 
and $r_Z$ to the quantity
\be
\left( \frac{g}{\sqrt{2}} |V_{tb}^*V_{cb}| \frac{m_b^2}{M_W^2} \right)^2   
\sim 10^{-8}   
\label{equa}
\ee 
(where $V_{ij}$ are the Kobayashi-Maskawa matrix elements \cite{CKM})
arising from 
the higher-order in the weak coupling and the GIM suppression 
mechanism of the one-loop decay width. 
The large discrepancy between the value of the ratio $r_H$ 
in eq.~(\ref{rhiggs}) and what is expected from the factor in 
eq.~(\ref{equa}), which on the other hand is supported by the value of 
$r_Z$, is a further indication that the values for $B(\tcht)$
reported in eq.~(\ref{sbagl}) can be incorrect.


In order to clarify the situation of the \tch decay,
we recomputed from scratch the complete 
analytical decay width for \tch (see ref.~\cite{soddu} for details
and analytical results). Our calculation has been done
putting $\mc=0$, which automatically selects in the 
final amplitude only the dominant contribution proportional to $(1+\gamma_5)$. 
The corresponding numerical results for $B(\tcht)$, when
$\mt = 175$GeV and $\Gt \simeq 1.55$ GeV, 
are reported in Table 1, in the column labeled 
by $(\mc=0)$, and in Fig.~3 (solid line). 
We used  $\mw=80.3$GeV, $\mb=5$GeV, $m_s=0.2$GeV,
and for the Kobayashi-Maskawa matrix elements
$|V_{tb}^*V_{cb}|=0.04$, according to  ref.~\cite{parti}.
Furthermore, we assumed 
$|V_{ts}^*V_{cs}|=|V_{tb}^*V_{cb}|$. As a consequence,
the $\md$ dependence in the amplitude drops out.

\noindent
Our results are several orders of magnitude smaller 
than the ones reported in ref.~\cite{sonid}.
In particular, for $\mh \simeq \mz$ we obtain
\be
B_{new}(\tcht) \simeq 1.2\cdot 10^{-13}  
\label{nostro}
\ee
to be compared with the corresponding value presented in ref.~\cite{sonid}
\be
B_{old}(\tcht) \simeq 6\cdot 10^{-8}.  
\label{loro}
\ee
The new value of $B(\tcht)$ in eq.~(\ref{nostro}) 
gives now $r_H \sim 4 \cdot 10^{-7}$.

In order to trace back the source of this inconsistency, we 
performed a thorough study of the analytical formula in eq.~(3)
of ref.~\cite{soniu}, for the decay width of the fourth-family
down-type quark $b'\to bH$, that is  the basis
for the numerical evaluation of $B(\tcht)$ presented in 
ref.~\cite{sonid}.

The results of this study can be summarized in the following way.
\begin{itemize}
\item
We recomputed and succeeded in reproducing the complete 
analytical expression for
$\Gamma(b'\to bH)$ in eq.~(3) of ref.~\cite{soniu}, 
including the relevant form factors $\alpha$ and $\beta$,
governing, respectively, the $(1+\gamma_5)$ and the 
$(1-\gamma_5)$  part of the amplitude.
\item
We find good agreement with all the numerical results on the decay
$b'\to bH$ presented
in ref.~\cite{soniu}, and based on eq.~(3) of the same reference.
\item
We performed the numerical evaluation\footnote{\noindent We 
checked this result  by computing the loop scalar integrals 
by both our own routines and the FF library Fortran routines \cite{fflib}.}
of $B(\tcht)$ by applying eq.~(3) 
in ref.~\cite{soniu} to the up-type quark decay \tchm. For $\mc=0$, 
we find complete agreement with our independent 
evaluation presented in the first column of Table 1.
For completeness, in the last column of Table 1, we  present
the effect of assuming in  $B(\tcht)$  a realistic value of 
the charm-quark mass $\mc$, according to eq.~(3) in \cite{soniu}. 
One can see that this effect
is smaller than about 1\permille  for $\mh\lta 150$GeV.
\item
The main result of the study is that our numerical calculation disagrees 
by either 5 or 6 orders of
magnitude (depending on $\mt$ and $\mh$) with the values of $B(\tcht)$ 
presented in Fig.~3 of ref.~\cite{sonid}. 
The latter are also based on eq.~(3) of ref.~\cite{soniu}.
In Fig.~4\footnote{\noindent The top total decay width has been
computed taking into account the complete $\mb$ effects \cite{total}.}, 
 we present the updated version of  Fig.~3 of
 ref.~\cite{sonid}.   
\end{itemize}

\noindent
A possible explanation for this situation could be some error in the computer 
code used by the authors of ref.~\cite{sonid} to work out their Fig.~3. 

It is interesting to apply the complete calculation 
for the  \tch width in the unrealistic parameters range
\be
M_W^2{\gg}m_t^2{\gg}m_b^2{\gg}m_c^2,    
\label{approx}
\ee
where it can be compared with the results of an approximated 
matrix element, that is valid in the same parameters range \cite{wille}
\be
{\cal M}(t\to cH)  \simeq  \frac{3g^3}{256\pi^2}V_{tb}^*V_{cb}    
\frac{m_tm_b^2}{M_W^3}\bar{u}_c(p_1)(1+\gamma^5)u_t(p_2).  
\label{eq:amptop}    
\ee
This simple expression shows the basic dependence of the amplitude on
both the top Yukawa coupling, through the factor $\mt/\mw$,
and the remnants of the GIM cancellation, giving rise to the factor
$V_{tb}^*V_{cb} \frac{\mbb}{\mww}$.
A straightforward evaluation of the \tch width according to 
eq.~(\ref{eq:amptop}) gives, for $\mt\simeq 30$GeV and $\mh\ll \mt$,
\be
\Gamma_{approx}(\tcht) \simeq  4.3 \cdot 10^{-16}   \GeV,
\label{appres}
\ee
to be compared with
\be
\Gamma_{exact}(\tcht) \simeq  6.7 \cdot 10^{-16}   \GeV.
\label{exares}
\ee
In the real case of $\mt\simeq 175$GeV,
there are two main limitations of the approximation in eq.~(\ref{eq:amptop}).
First, the amplitude in eq.~(\ref{eq:amptop}) does not take into 
account the absorbitive
contributions  arising when $\mt \gta \mw +\mb$.
Secondly,  it neglects the terms of order $\mtt/\mww$ 
(note that both the terms of order $\mt^3/\mw^3$ and the 
ultraviolet singularities  are cancelled by the GIM mechanism).


In Fig.~5,  we compare as a function of $\mt$
the behaviour of our calculation for $\Gamma(\tcht)$ (solid line)
with the approximated one obtained through eq.~(\ref{eq:amptop})
(dashed line).
Here, we assume the unphysical value
$\mh=1$GeV, in order to
 be able to explore the approximation
down to very low $\mt$, maintaining  the parameters
in its validity region. 
It is worth noting that the simple amplitude 
of eq.~(\ref{eq:amptop}) reproduces very well the exact behaviour of
$\Gamma(\tcht)$ for $\mt\lta 40$GeV. Note also 
the peak produced in the complete calculation
by the opening of the
absorbitive contribution for $\mt \gta \mw+\mb$,
that is obviously absent in the approximated estimate.
In Fig.~3, we also display, for the sake of comparison,
 the behaviour for $B(\tcht)$
calculated through the approximated formula eq.~(\ref{eq:amptop})
and extrapolated out of its validity region, at the fixed
value $\mt=175$GeV, as a function of $\mh$. 
For  values of the 
Higgs mass not excluded experimentally
(presently $\mh\gta 89$GeV \cite{mhig}),
the approximated amplitude gives results differing from the exact
ones by less than a factor 4. The change in the slope of the exact 
result for $\mh\simeq 160$GeV is the effect of
the  threshold for a new absorbitive contribution
to the amplitude for $\mh\gta 2\mw$.

In conclusion, in this letter we have  pointed out that the widely quoted
result of ref.~\cite{sonid} establishing a relatively large branching 
ratio for the decay
\tch in the SM has been  overestimated by several orders of magnitude.
On the other hand, we agree with the analytical formulae in \cite{soniu},
by which this result was worked out.
The corrected numerical estimates are shown in Table 1.
We find $B(\tcht)\simeq 1 \cdot 10^{-13}\div 4\cdot 10^{-15}$
for $\mz \lta \mh \lta 2\mw$, that is about 6 orders of magnitude
less than previously evaluated, and usually quoted in the literature.
Such a small rate will not be measurable even at the highest
luminosity accelerators that are presently conceivable.
An eventual experimental signal in the rare $t$ decays will definitely 
have to be ascribed to some new physics effect.

\vskip 0.3cm
\noindent
We wish to warmly thank V.A.~Ilyin for discussions and suggestions.
\vskip 0.7cm
After submission of this paper and appearance of the preprint
in the hep-ph bulletin, we received a message by one of the authors of 
ref.~\cite{sonid} (J.L.H.), saying that they were aware of the error
in $B(\tcht)$, which indeed arose from a mistake in the numerical code. 
Unfortunately, a proper erratum never appeared. The correct
numbers (with which we agree) have been shown in Fig. 21 of ref.\cite{erra}, 
where, however, the disagreement with the previous values in
ref.~\cite{sonid} has not been stressed.

\newpage

\newpage
\begin{table}[t]  
\begin{center}  
\begin{tabular}{|c|c|c|} \hline   
		 &                     &                       \\   
$ m_H\;(GeV) $ & $ B(\tcht) \;(m_c=0) $ & $ B(\tcht) \;(m_c=1.5GeV) $ \\    
[2.mm]\hline\hline  
	&                           &                       \\    
  60  & $ 0.2557\cdot10^{-12} $ & $ 0.2558\cdot10^{-12} $ \\ [2.mm]\hline   
	&                           &                       \\   
  70  & $ 0.1986\cdot10^{-12} $ & $ 0.1986\cdot10^{-12} $ \\ [2.mm]\hline   
	&                           &                       \\   
  80  & $ 0.1532\cdot10^{-12} $ & $ 0.1532\cdot10^{-12} $ \\ [2.mm]\hline   
	&                           &                       \\   
  90  & $ 0.1169\cdot10^{-12} $ & $ 0.1169\cdot10^{-12} $ \\ [2.mm]\hline   
	&                           &                       \\   
  100 & $ 0.8775\cdot10^{-13} $ & $ 0.8777\cdot10^{-13} $ \\ [2.mm]\hline   
	&                           &                       \\   
  110 & $ 0.6451\cdot10^{-13} $ & $ 0.6452\cdot10^{-13} $ \\ [2.mm]\hline   
	&                           &                       \\   
  120 & $ 0.4605\cdot10^{-13} $ & $ 0.4605\cdot10^{-13} $ \\ [2.mm]\hline   
	&                           &                       \\   
  130 & $ 0.3146\cdot10^{-13} $ & $ 0.3146\cdot10^{-13} $ \\ [2.mm]\hline   
	&                           &                       \\   
  140 & $ 0.1999\cdot10^{-13} $ & $ 0.1998\cdot10^{-13} $ \\ [2.mm]\hline   
	&                           &                       \\   
  150 & $ 0.1107\cdot10^{-13} $ & $ 0.1105\cdot10^{-13} $ \\ [2.mm]\hline   
	&                           &                       \\   
  160 & $ 0.4424\cdot10^{-14} $ & $ 0.4410\cdot10^{-14} $ \\    
	&                           &                       \\   
  [2.mm]\hline\hline   
\end{tabular}   
\vspace*{1.4cm}
\caption{ Branching ratio for the decay 
\tch versus $\mh$ for both  massless and massive charm quark. 
In the column ($m_c=0$), we show the values obtained by our own calculation,
in the column  ($m_c=1.5$GeV) we give, for comparison, the values obtained by
our numerical evaluation of the formulae in ref.~\protect\cite{soniu}.
We assume $\mt=175$GeV.
}
\end{center}  
\end{table}    
\newpage
\begin{figure*}[t]
\begin{center}
\vspace*{-4.cm}
\mbox{\epsfxsize=16cm\epsfysize=18.5cm\epsffile{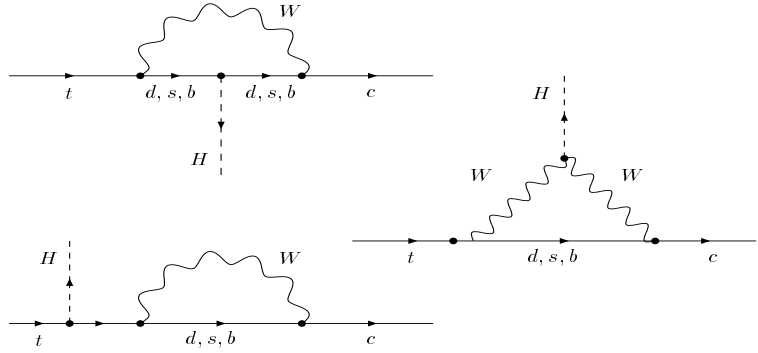}}
\vspace*{-6.5cm}
\caption{ Feynman graphs for the decay \tch in the unitary gauge
($\mc=0$ is assumed).
 }
\end{center}
\end{figure*}
\begin{figure*}[t]
\begin{center}
\vspace*{-2.cm}
\mbox{\epsfxsize=16cm\epsfysize=18.5cm\epsffile{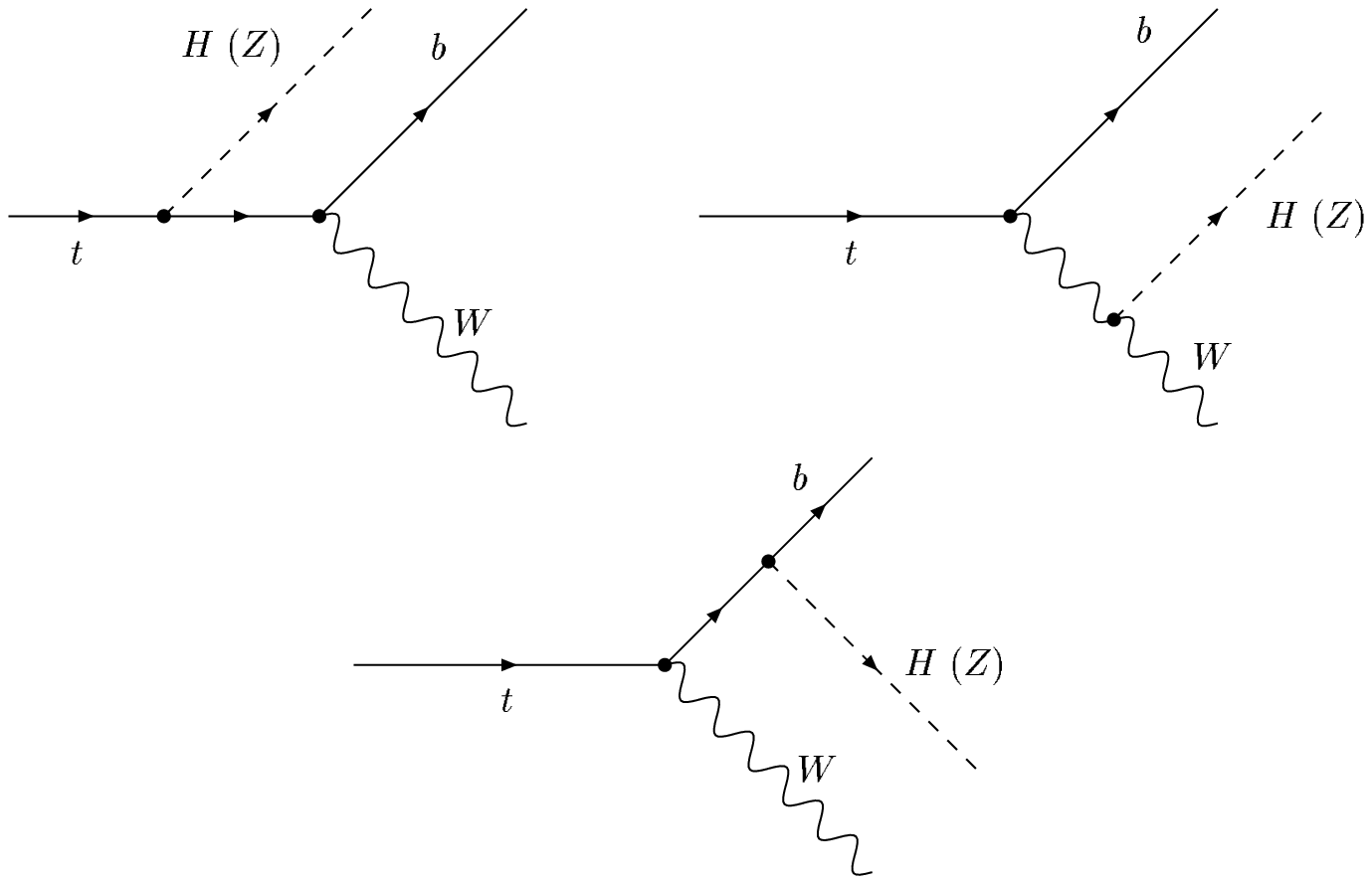}}
\vspace*{-7.5cm}
\caption{  Feynman graphs for the decay \tbwh (\tbwzm).
 }
\end{center}
\end{figure*}
\begin{figure*}[htbp]
\begin{center}
\mbox{\epsfxsize=18cm\epsfysize=22cm\epsffile{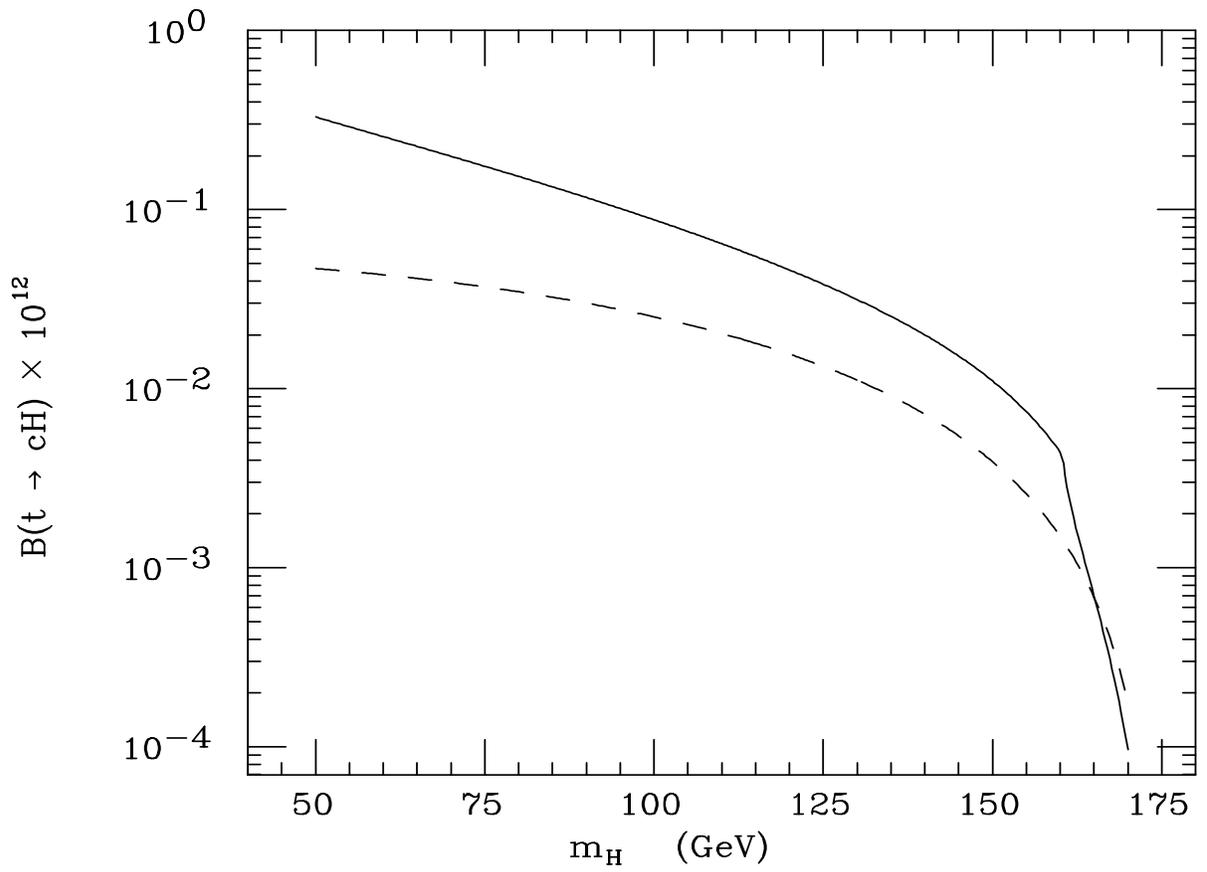}}
\vspace*{-4.cm}
\caption{ Branching ratio for the decay \tch versus
 $\mh$, for $\mt=175$GeV and $\Gt=1.55$GeV.
The solid line is the result of our calculation. The dashed line shows 
the approximated behaviour
according to eq.~(\protect\ref{eq:amptop}).
 }
\end{center}
\end{figure*}
\begin{figure*}[t]
\begin{center}
\mbox{\epsfxsize=18cm\epsfysize=22cm\epsffile{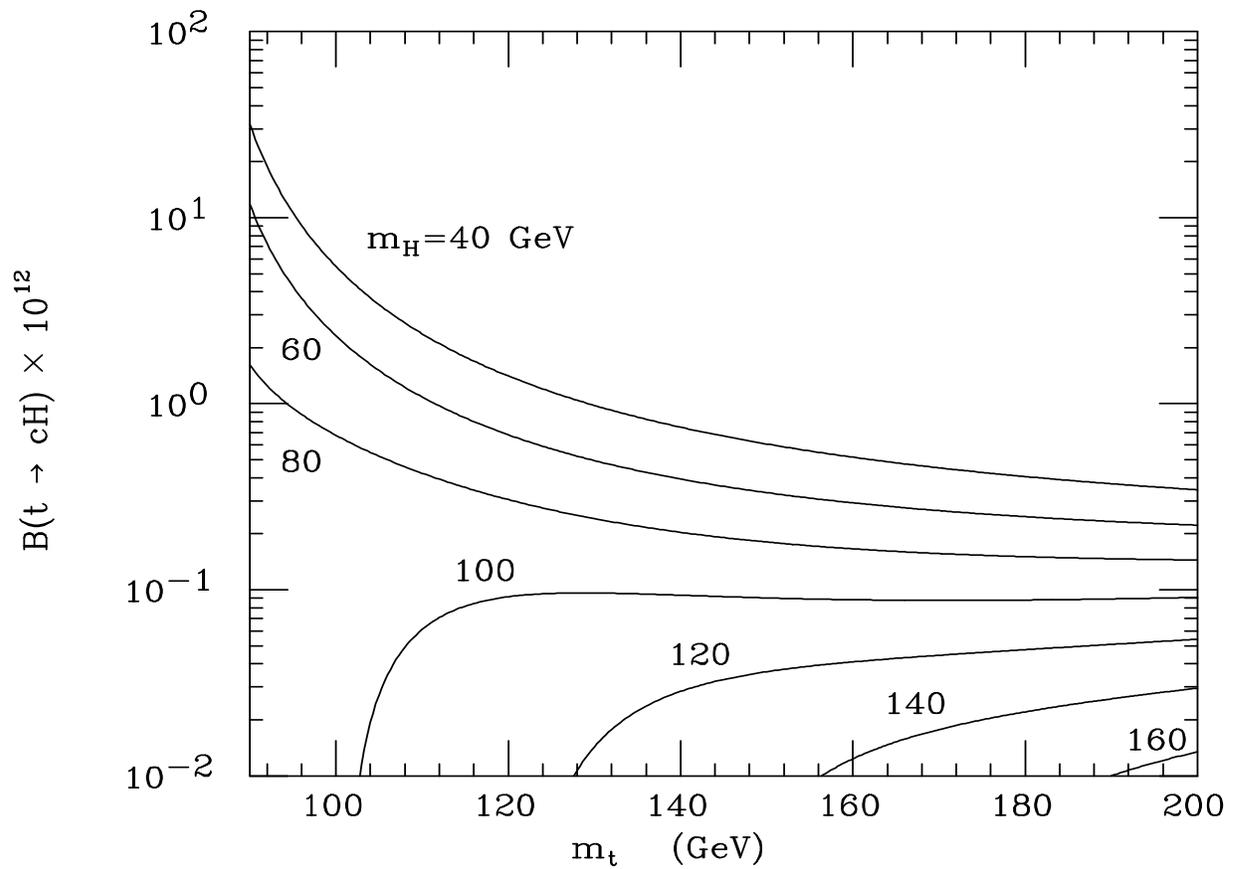}}
\vspace*{-4.cm}
\caption{  Branching ratio for the decay \tch 
as a function of $\mt$ for different values of $\mh$.
 }
\end{center}
\end{figure*}
\begin{figure*}[b]
\begin{center}
\mbox{\epsfxsize=18cm\epsfysize=22cm\epsffile{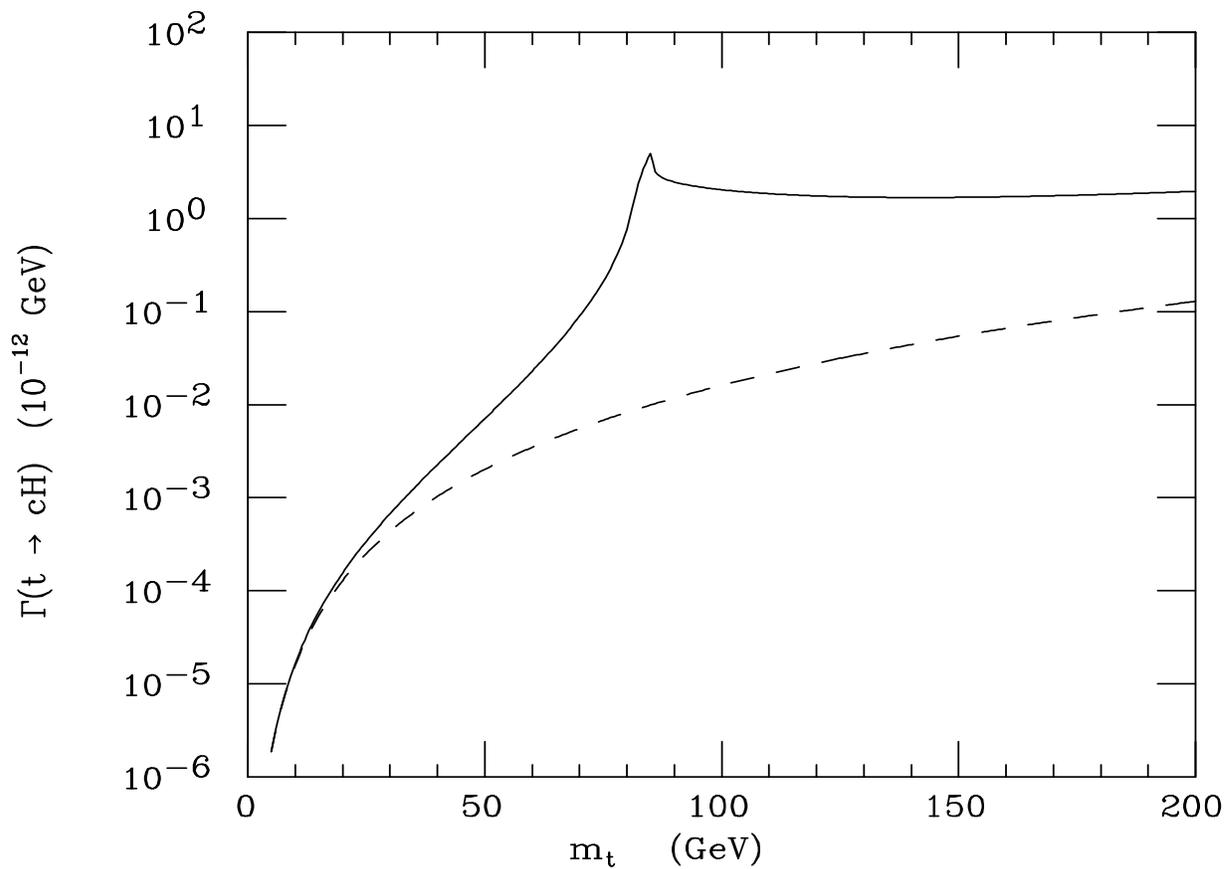}}
\vspace*{-4.cm}
\caption{Comparison between the widths of \tchm, as functions
of $\mt$, at fixed (unrealistic) $\mh=1$GeV, and $\mc=0$,
calculated through our complete
formulae (solid line) and the approximate expression in 
eq.~(\protect\ref{eq:amptop}) 
(dashed line). 
}
\end{center}
\end{figure*}
\end{document}